\documentclass[pra,a4paper,showpacs,twocolumn,superscriptaddress,longbibliography]{revtex4-1}

\usepackage{amssymb}
\usepackage{amsmath}
\usepackage{amsfonts}
\usepackage{graphicx}
\usepackage{bm}
\usepackage{color}
\usepackage{multirow}
\usepackage{natbib}
\usepackage{hyperref}

\DeclareMathOperator{\Tr}{Tr}
\DeclareMathOperator{\spp}{sp}
\DeclareMathOperator{\sgn}{sgn}

\DeclareMathOperator{\im}{Im}
\DeclareMathOperator{\re}{Re}

\begin{document}

\title{Quantum corrections to conductivity of disordered electrons  due to inelastic scattering off magnetic impurities}

\author{I. S.~Burmistrov}

\affiliation{L.D. Landau Institute for Theoretical Physics, Kosygina
  street 2, 117940 Moscow, Russia}
  
\affiliation{Moscow Institute of Physics and Technology, 141700 Moscow, Russia}

\affiliation{Institut f\"ur Theorie der kondensierten Materie, Karlsruhe Institute of Technology, 76128 Karlsruhe, Germany}

\affiliation{Institut f\"ur Nanotechnologie, Karlsruhe Institute of Technology, 76021 Karlsruhe, Germany}

\author{E. V. Repin}

\affiliation{Condensed-matter Physics Laboratory, National Research University Higher School of Economics, 101000 Moscow, Russia}

\affiliation{Delft University of Technology, The Netherlands}

\begin{abstract}
We study the quantum corrections to the conductivity of the two-dimensional disordered interacting electron system in the diffusive regime due to inelastic scattering off rare magnetic impurities. We focus on the case of very different g-factors for electrons and magnetic impurities. Within the Born approximation for the inelastic scattering off magnetic impurities we find additional temperature-dependent corrections to the conductivity of the Altshuler-Aronov type. Our results demonstrate that the low temperature transport in interacting disordered electron systems with rare magnetic impurities is more interesting than it was commonly believed on the basis of treatment of magnetic impurity spins as classical ones.
\end{abstract}

\maketitle

\section{Introduction}

As it is well known, the low temperature properties of an electron system are significantly affected by electron scattering off rare magnetic impurities. The simplest approach is to treat a magnetic impurity classically as a random three-dimensional vector of a fixed length. Despite that such model ignores a quantum dynamics of the spin, i.e. treats the scattering off a magnetic impurity 
elastically, the model is powerful enough to produce a number of interesting, nontrivial effects, e.g. suppression of the superconducting transition temperature due to elastic electron spin-flip \cite{AG}, suppression of temperature dependence of the weak localization correction  to conductivity \cite{Lee1980,Efetov1980b}, etc.

The quantum dynamics of the spin of a magnetic impurity is responsible for the Kondo effect: renormalization of the exchange coupling between an electron and impurity spins that leads to non-monotonic temperature dependence of resistivity  \cite{Kondo}.
Physically, the quantum dynamics of the spin allows an electron to scatter inelastically off a magnetic impurity \cite{Zarand2004,Garst2005,Borda2007}. For example, the Zeeman splitting of the magnetic impurity levels results in energy dependence of the spin-flip scattering even within the Born approximation \cite{Vavilov2003}. The presence of potential elastic scattering together with the inelastic spin-flip scattering results in modification of the Kondo effect and the behaviour of the quantum corrections to conductivity \cite{Ohkawa1983,Vladar1985,Kettemann2007}. 
For example, in disordered electron systems the inelastic spin-flip scattering affects the weak localization correction and mesoscopic conductance fluctuations via the energy-dependent dephasing time induced by spin-flip scattering \cite{Vavilov2003,Kettemann2006,Micklitz2006,Micklitz2007,Kashuba2016}.   In addition to the influence on the weak localization correction  the inelastic scattering off magnetic impurities results in appearance of the Altshuler-Aronov-type corrections to the conductivity \cite{Ohkawa1983,Ohkawa1984,Suga1986}. These temperature-dependent corrections  have been found in the third order in the exchange interaction. It can be easily argued why this is the lowest order in which such corrections can arise. Indeed, in order to have inelastic scattering off a magnetic impurity within the Born approximation one needs to have the Zeeman splitting. However, the Zeeman splitting induces a cut-off for the relevant diffusive modes. This forbids temperature-dependent corrections to conductivity in the second order in the exchange interaction. 
However, the above arguments assume that the Zeeman splitting for a magnetic impurity and for an electron are the same which is true provided the L\'ande factors are the same.

In this paper, we consider the case of very different  g-factors 
of an electron, $g_{\rm e}$, and a magnetic impurity, $g_{\rm i}$: $|g_{\rm i}| \gg |g_{\rm e}|$. In this case the impurity Zeeman splitting, $b_{\rm i} =g_{\rm i}\mu_B H$ can be much larger than the electron Zeeman splitting, $b_{\rm e}=g_{\rm e}\mu_B H$. Here $\mu_B$ stands for the Bohr magneton and $H$ denotes the external magnetic field. For a sake of concreteness we consider a two-dimensional electron system in parallel magnetic field $H$. Then, as we shall demonstrate, there exists the logarithmic-in-$T$ correction to the conductivity due inelastic scattering off magnetic impurities within the Born approximation provided the temperature satisfies inequalities:  
\begin{equation}
|b_{\rm e}| \ll T \ll |b_{\rm i}| \ . 
\label{eq:ineq}
\end{equation}
Also, we study how inelastic scattering off magnetic impurities interferes with the electron-electron interaction. We find that, on the one hand, the inelastic scattering off magnetic impurities modifies the Altshuler-Aronov correction, and, on the other hand, the electron-electron interaction affects the correction to the conductivity due to the inelastic scattering off magnetic impurities (which also exists in the absence of electron-electron interaction).

The paper is organized as follows. In Sec. \ref{Sec:Form} we remind the formalism of the Finkel'stein nonlinear sigma model. 
The perturbative expansion of the nonlinear sigma model and the structure of diffusive modes are discussed in Sec. \ref{Sec:Pert}. In Sec. \ref{Sec:Cond} we present our results for the temperature dependence of conductivity in two-dimensional electron system. We conclude the paper with the discussion of our findings (Sec. \ref{Sec:Sum}).

\section{Finkel'stein nonlinear sigma model \label{Sec:Form}}

We consider a two-dimensional interacting electron system in the presence of short-ranged potential disorder. In addition, we assume the presence of weak spin-flip scattering due to an exchange interaction between rare magnetic impurities and electrons  described by the following Hamiltonian
\begin{equation}
{H}_{\rm mag} = J \sum_{j} \psi^\dag(\bm{r}_j) \bm{S}_j \bm{\sigma} \psi(\bm{r}_j) \ .
\end{equation}
Here $\bm{\sigma}$ and $\bm{S}_j$ stand for the Pauli matrices and the spin operator of a magnetic impurity at the position $\bm{r}_j$, respectively. The electron creation and annihilation operators are denoted as $\psi^\dag(\bm{r})$ and $\psi(\bm{r})$.
We shall treat rare magnetic disorder under the following assumptions:
(i) impurity positions $\bm{r}_j$ have the Poisson distribution; 
(ii) impurity spins $\bm{S}_j$ are independent but have its own quantum dynamics. 

In the absence of magnetic scattering, the effective field theory 
for disordered interacting electrons in the diffusive regime, $T\ll 1/\tau$, where $\tau$ denotes the elastic mean free time,
is defined in a standard way (for a review see \cite{Finkelstein1990,BelitzKirkpatrick1994}). 
In the absence of magnetic field and magnetic impurities the Hamiltonian of the system preserves spin-rotational and time-reversal symmetries. Then the effective field theory is formulated in terms of a matrix field $Q \in G/K$ with $G={\rm Sp}(2N)$ and $K={\rm Sp}(N)\times {\rm Sp}(N)$. The rank of $G$ is given by $N= 4 N_r N_m$ where $N_m$ denotes the number of Matsubara frequencies involved and $N_r$ stands for the number of replica. For computation of physical observables one needs to take two limits: $N_m \to \infty$ and $N_r\to 0$, at the end of calculations. We note that the limit  $N_m \to \infty$ should be taken in a way consistent with the gauge invariance (see Ref. \cite{Baranov1999a} for details). The factor $4$ appears since one needs to take into account the spin and Nambu (particle-hole) spaces. Taken into account Zeeman splitting due to external magnetic field the effective action can be written as follows \cite{Finkelstein1983a,Finkelstein1983b,Finkelstein1984}
\begin{align}
\mathcal{S}_\sigma=& - \int d\bm{r} \Tr \Bigl [ \frac{g}{32}   (\nabla Q)^2 - 4\pi T Z_\omega  \eta Q - i Z_s b_{
\rm e} t_{33} Q \Bigr ] \notag \\
& - \frac{\pi T}{4} \sum_{\alpha,n,r,j}\Gamma_j 
\int d\bm{r} \Tr I_n^\alpha t_{rj} Q  \Tr I_{-n}^\alpha t_{rj} Q .
\label{eq:NLSM}
\end{align}
%The partition function can be computed as $\int \mathcal{D}[Q] \exp(\mathcal{S})$
Here the sixteen matrices $t_{rj}$, $j,r=0,1,2,3$ act 
in a tensor product of the spin (subscript $j$) and Nambu  (subscript $r$) spaces:
% The indices $r$ and $j$ of matrices $t_{rj}$ indicate its decomposition by the basis of tensor products of the unit and Pauli matrices in these spaces, i.e. 
\begin{equation}
\label{trj}
t_{rj} = \tau_r\otimes s_j, \qquad r,j = 0,1,2,3  .
\end{equation}
Here matrices $\tau_0$ and $s_0$ stand for the $2\times 2$ unit matrices and
\begin{equation}
\tau_1/s_1  = \begin{pmatrix}
0 & 1\\
1 & 0
\end{pmatrix}, \, \tau_2/s_2 = \begin{pmatrix}
0 & -i\\
i & 0
\end{pmatrix}, \, \tau_3/s_3 = \begin{pmatrix}
1 & 0\\
0 & -1
\end{pmatrix} . \notag
\end{equation}
The effective action \eqref{eq:NLSM} involves the following matrices
\begin{equation}
%\Lambda_{nm}^{\alpha\beta} = \sgn n \, \delta_{nm} \delta^{\alpha\beta}t_{00}, \quad
\eta_{nm}^{\alpha\beta}=n \, \delta_{nm}\delta^{\alpha\beta} t_{00},  \quad
(I_k^\gamma)_{nm}^{\alpha\beta}=\delta_{n-m,k}\delta^{\alpha\beta}\delta^{\alpha\gamma} t_{00} ,
\end{equation}
where $\alpha,\beta = 1,\dots, N_r$ stands for replica indices and indices $n,m$ correspond to the
Matsubara fermionic frequencies $\varepsilon_n = \pi T (2n+1)$. 
The total (including spin) dimensionless  (in units $e^2/h$) Drude conductivity is denoted by $g$. The interaction amplitudes $\Gamma_j$ (for the singlet channel, $\Gamma_0 = \Gamma_s$, and for the triplet channel, $\Gamma_1=\Gamma_2=\Gamma_3=\Gamma_t$) describe electron-electron interaction in the particle-hole channel. 
In what follows it will be also convenient to use $\gamma_j = \Gamma_j/Z_\omega$ and $\gamma_{s,t}=\Gamma_{s,t}/Z_\omega$.
The parameter $Z_\omega$ takes into account nontrivial frequency renormalization under the renormalization group \cite{Finkelstein1983a,Finkelstein1983b,Finkelstein1984}. We note that the bare value of the parameter $Z_\omega$ is equal to $\pi \nu/4$ where $\nu$ denotes the density of states at the Fermi level.
The last term in the first line of Eq. \eqref{eq:NLSM} describes the effect of the parallel magnetic field on electrons. This term violates explicitly time-reversal symmetry. The quantity $Z_s = Z_\omega+\Gamma_t$ describes the Femi-liquid-type enhancement of the g-factor (see reviews \cite{Finkelstein1990,BelitzKirkpatrick1994} for details). We note that in this paper we neglect the electron-electron interaction in the Cooper channel. 

%This allows us to neglect the effect of parallel magnetic field on transport of electrons. However, the presence of large Zeeman splitting will be crucial for dynamics of magnetic impurities. 

By construction the matrix $Q(\bm{r})$ describes local rotations around the spatially independent matrix $\Lambda$:
\begin{equation}
Q=\mathcal{T}^{-1} \Lambda \mathcal{T}, \quad \Lambda_{nm}^{\alpha\beta} = \sgn \varepsilon_n \, \delta_{nm} \delta^{\alpha\beta}t_{00} .
\label{eq:QQ}
\end{equation}
Here the matrices $\mathcal{T} \in G$ obey the following symmetry relations:
\begin{equation}
C (\mathcal{T}^{-1})^{\rm T} = \mathcal{T} C,\qquad \mathcal{T}^{\rm T} C = C \mathcal{T}^{-1} , \label{TC}
 \end{equation}
where $C = i t_{12}$. The symbol $\mathcal{T}^{\rm T}$ denotes the matrix transpose of $\mathcal{T}$.  As the consequence of Eqs. \eqref{eq:QQ} and \eqref{TC}, the matrix $Q$ is subjected to the local nonlinear constraint, $Q^2(\bm{r})=1$, satisfies the condition, $\Tr Q=0$, and obeys charge-conjugate relation,
\begin{equation}
 Q=Q^\dag = C^T Q^T C .
 \label{TC2}
\end{equation}

In the presence of magnetic impurities the full effective action $\mathcal{S}$ is the sum of the Finkel'stein nonlinear sigma model $\mathcal{S}_\sigma$ and the additional part $\mathcal S_\text{mag}$, i.e. $\mathcal{S}=\mathcal{S}_\sigma+\mathcal 
S_\text{mag}$. For rare magnetic impurities the latter can be written as a sum over contributions of individual magnetic impurities \cite{MaS}:
\begin{equation}
\mathcal S_\textrm{mag}  = \frac 12 \sum_j \Tr \ln \left( 1+ i\pi \nu J \, Q(\bm{r}_j) \tau_3 \bm{\sigma} \hat{\bm{S}}_{j} \right) .
\label{Ssep}
\end{equation}
Here we introduce the following notations:
\begin{equation}
\hat{\bm{S}}_j = \sum_n {\bm{S}}_j(i\omega_n) I_n ,\quad 
{\bm{S}}_j(i\omega_n) = \int\limits_0^\beta d\tau {\bm{S}}_j(\tau)e^{i\omega_n\tau} , 
\end{equation}
where $\beta = 1/T$, $\omega_n=2\pi T n$, and the matrix $I_n$ is defined as follows
\begin{equation}
(I_k)_{nm}^{\alpha\beta}=\delta_{n-m,k}\delta^{\alpha\beta} t_{00} .
\end{equation}
We note that the form \eqref{Ssep} of the action $\mathcal S_\text{mag}$ is equivalent to the self-consistent $T$-matrix approximation for magnetic scattering, i.e. it is derived by taking into account all orders in scattering off a single magnetic impurity but by neglecting contributions with intersecting impurity lines. 

We perform the Poisson averaging over positions of the magnetic impurities with the help of the following relation~\cite{Friedberg1975}
\begin{equation}
\Bigl \langle \exp \sum_j f(\bm{r_j}) \Bigr \rangle = \exp \left \{ n_s \int d\bm{r} \Bigl [e^{f(\bm{r})}-1\Bigr ]\right \} ,
\end{equation}
where $n_s$ denotes the average concentration of magnetic impurities. Then we find that the contribution to the effective action due to magnetic impurities becomes
\begin{equation}
\mathcal S_\text{mag} \to
    {n}_s \int d \bm{r}
  \left ( \left \langle e^{\frac 12 \Tr \ln \left( 1+ i\pi \nu J \, Q(\bm{r})\tau_3 \bm{\sigma} \hat{\bm{S}} \right)}\right \rangle_{\bm{S}}
  -1 \right) .
\label{Smagn0}
\end{equation}
Here $\langle \dots \rangle_{\bm{S}}$ stands for the averaging over dynamics of a single magnetic impurity.

In this paper we restrict our consideration by the Born approximation for the scattering off a single magnetic impurity. 
Therefore, we can expand $\Tr \ln$ in Eq. \eqref{Smagn0} upto the second order in $J$. Then we find
\begin{gather}
\mathcal S_\text{mag}  = \frac{n_s\pi \nu J}{2} \int d\bm{r} 
\Bigl \langle  i \Tr Q \tau_3 \bm{\sigma} \langle \hat{\bm{S}}\rangle_{\bm{S}} 
+ \frac{\pi \nu J}{2}  \Tr \bigl (Q \tau_3 \bm{\sigma}\hat{\bm{S}}\bigr )^2
\notag \\
 -
\frac{\pi \nu J}{4}  \Bigl (\Tr Q \tau_3 \bm{\sigma}\hat{\bm{S}}\Bigr )^2 \Bigr \rangle_{\bm{S}} .
\end{gather}

In order to proceed further we need to perform averaging over dynamics of the spin of a magnetic impurity in $\mathcal S_\text{mag}$. Neglecting a back action of electrons on the spin of a magnetic impurity, allows us to write the impurity Hamiltonian as follows: $H_\textrm{i} = b_{\rm i} S_z$. Then, we need the corresponding Matsubara spin-spin correlation functions:
\begin{equation}
\chi_{\pm}(\tau_1,\tau_2) = \frac{1}{S(S+1)} \begin{cases}
\langle S_\pm(\tau_1) S_\mp(\tau_2)\rangle_{\bm S} , &
\, \tau_1>\tau_2 ,\\
\langle S_\mp(\tau_2) S_\pm(\tau_1)\rangle_{\bm S} , &
\, \tau_2>\tau_1 ,
\end{cases}
\end{equation}
where $S_\pm=S_x\pm i S_y$, and 
\begin{equation}
\chi_{zz}(\tau_1,\tau_2) =\frac{1}{S(S+1)}\begin{cases}
\langle S_z(\tau_1) S_z(\tau_2)\rangle_{\bm S} , &
\, \tau_1>\tau_2 ,\\
\langle S_z(\tau_2) S_z(\tau_1)\rangle_{\bm S} , &
\, \tau_2>\tau_1 .
\end{cases}
\end{equation}
Using the equations of motion for a free spin in a magnetic field we 
find the following results:
\begin{equation}
\chi_{\pm}(i\omega_n)  =   - e^{i\omega_n 0^+} \frac{2 M_1}{i\omega_n\pm b_{\rm i}}
, \quad
\chi_{zz}(i\omega_n)   =  \delta_{n,0}\beta  M_2 \ .
\label{eq:SpinSusc}
\end{equation}
Here we introduced
\begin{equation} 
M_n = \frac{1}{S(S+1)}\sum_{m=-S}^{m=S} m^n e^{-\beta b_{\rm i}m} \Biggl / \Biggr .\sum_{m=-S}^{m=S}  e^{-\beta b_{\rm i}m} .\end{equation}
We note the following useful relations $e^{-\beta b_{\rm i}} 
\langle S_- S_+ \rangle_{\bm S} = \langle S_+ S_- \rangle_{\bm S}$ and $M_2 = 1 + M_1 \coth(b_{\rm i}/2T)$. Using the results \eqref{eq:SpinSusc}, we obtain
\begin{align}
\mathcal S_\text{mag}  = & \int d\bm{r}  \Biggl \{ \frac{i}{2} n_s \pi \nu J  \langle S_z\rangle_{\bm{S}} 
\Tr t_{33} Q
+ \frac{Z_\omega T}{4\tau_{s0}}
 \sum_{n} \chi_+(i\omega_n)
 \notag \\
& \times
  \Bigl [ \Tr 
t_{-} I_n Q
t_{+} I_{-n} Q
 -
\frac{1}{2}  \Tr t_{-} I_n Q \Tr t_{+} I_{-n} 
Q \Bigr ] 
\notag \\
& + \frac{Z_\omega}{2\tau_{s0}}M_2 
 \Bigl [ \Tr 
t_{33}   Q
t_{33}  Q
 -
\frac{1}{2}  \Tr t_{33}   Q \Tr t_{33}  Q
\Bigr ]  \Biggr \}
 ,
\label{eq:Smag}
\end{align}
where $t_{\pm}=t_{31}\pm it_{32}$ and 
\begin{equation}
\frac{1}{\tau_{s0}} = \frac{n_s(\pi \nu J)^2S(S+1)}{2Z_\omega} 
\end{equation}
denotes the classical spin-flip rate at zero magnetic field.

The first term in the right hand side of Eq. \eqref{eq:Smag} 
corresponds to additional Zeeman splitting of electrons due to magnetization of magnetic impurities. The second term in the right hand side of Eq. \eqref{eq:Smag}  describes the contribution due to inelastic spin-flip scattering off magnetic impurity. We emphasize that contrary to the term due to electron-electron interaction, see the second line in Eq.
\eqref{eq:NLSM}, the inelastic term due to scattering off magnetic impurities mixes different replica channels.

\section{Perturbative expansion\label{Sec:Pert}}

For the perturbative treatment (in $1/g$) of the action $\mathcal{S}_\sigma+\mathcal{S}_\textrm{mag}$
we  need to resolve the constraint $Q^2=1$. In order to do it we 
use the square-root parametrization:
\begin{gather}
Q = W +\Lambda \sqrt{1-W^2}\ , \qquad W= \begin{pmatrix}
0 & w\\
\bar{w} & 0
\end{pmatrix} .
\label{eq:Q-W}
\end{gather}
In what follows we shall adopt the following notations: $W_{n_1n_2} = w_{n_1n_2}$ and $W_{n_4n_3} = \bar{w}_{n_4n_3}$ where $n_{1,3}\geqslant 0$ and $n_{2,4}<0$. The two blocks of the matrix $W$  are related by the following symmetry relation as
\begin{gather}
\bar{w} = -C w^T C .
\end{gather}
We note that here the matrix transposition acts on the Matsubara space indices. Expansion of $\mathcal{S}_\sigma+\mathcal{S}_\textrm{mag}$  to the second order in  $W$ yields the following Gaussian action:
\begin{widetext}
\begin{gather}
\mathcal{S}_\sigma^{(2)}+\mathcal{S}_\textrm{mag}^{(2)} =
- 4 \int \frac{d\bm{p}}{(2\pi)^d}\sum_{rr^\prime;jj^\prime} \sum_{\alpha_{l},n_l}
\bigl [w_{rj}(\bm{p})\bigr ]_{n_1n_2}^{\alpha_1\alpha_2}
\bigl [\bar{w}_{r^\prime j^\prime}(-\bm{p})\bigr ]_{n_4n_3}^{\alpha_4\alpha_3} \delta_{n_{12},n_{34}} 
\Biggl \{
\delta_{n_1n_3}\delta_{n_2n_4} \delta^{\alpha_1\alpha_3}
\delta^{\alpha_2\alpha_4} 
\Biggr [
\delta_{jj^\prime}\delta_{rr^\prime}
Z_\omega
 \Biggl ( D p^2  
\notag \\
+ \Omega^\varepsilon_{12}+ \frac{1}{\tau_{rj}^{\rm sf}}
+ \frac{1}{\tau_{\perp}^{\rm sf}} \Bigl (
h(i\varepsilon_{n_1}) + h(-i\varepsilon_{n_2})
\Bigr )
\Biggr )
- Z_s \tilde{b}_{\rm e}
\bigl (\delta_{r0}\delta_{r^\prime 3}+\delta_{r3}\delta_{r^\prime 0} \bigr ) \mu_{jj^\prime}^{(d)}
- Z_s \tilde{b}_{\rm e}
\bigl (\delta_{r1}\delta_{r^\prime 2}-\delta_{r2}\delta_{r^\prime 1} \bigr ) \mu_{jj^\prime}^{(c)}
\Biggl ]
 %+ \frac{n_s(\pi\nu J)^2 T}{2Z_\omega} \delta_{rr^\prime} \delta_{n_1n_2}\delta_{m_1m_2} \delta^{\alpha_1\alpha_2} \delta^{\beta_1\beta_2} \big (\delta_{j0}\delta_{j^\prime 3} + \delta_{j3}\delta_{j^\prime 0} \bigr ) i \sum_{k\neq 0} \im \chi_\perp(i\omega_k) \bigl[\theta(n_1-k)-\theta(m_1-k)\bigr ]
%\notag \\
%% vanish due to the symmetry relation
%%\bigl [w_{r0}(\bm{p})\bigr ]_{n_1m_1}^{\alpha_1\beta_1} \bigl [\bar{w}_{r 3}(-\bm{p})\bigr ]_{m_1n_1} ^{\beta_1\alpha_1} =  - \bigl [w_{r3}(\bm{p})\bigr ]_{n_1m_1}^{\alpha_1\beta_1} \bigl [\bar{w}_{r 0}(-\bm{p})\bigr ]_{m_1n_1} ^{\beta_1\alpha_1} 
%%%%%%%%%%%%
%+\frac{n_s(\pi\nu J)^2 T}{2Z_\omega} \delta_{rr^\prime} \delta_{n_1n_2}\delta_{m_1m_2} \delta^{\alpha_1\alpha_2} \delta^{\beta_1\beta_2} \big (\delta_{j1}\delta_{j^\prime 2} - \delta_{j2}\delta_{j^\prime 1} \bigr ) \sum_{k\neq 0} \im \chi_\perp(i\omega_k) \bigl[\theta(n_2-k)-\theta(k-m_1)\bigr ] \notag \\
%% vanish due to the symmetry relation
%%\bigl [w_{r1}(\bm{p})\bigr ]_{n_1m_1}^{\alpha_1\beta_1} \bigl [\bar{w}_{r2}(-\bm{p})\bigr ]_{m_1n_1} ^{\beta_1\alpha_1} =   \bigl [w_{r2}(\bm{p})\bigr ]_{n_1m_1}^{\alpha_1\beta_1} \bigl [\bar{w}_{r1}(-\bm{p})\bigr ]_{m_1n_1} ^{\beta_1\alpha_1} 
%
%
\notag \\
%+ \frac{2 n_s(\pi\nu J)^2 T}{4Z_\omega} \delta_{n_{12},m_{12}} \delta_{rr^\prime} \delta^{\alpha_1\beta_1} \delta^{\alpha_2\beta_2} \delta_{r3}\bigl (\delta_{j1}\delta_{j^\prime2}-\delta_{j2}\delta_{j^\prime 1}\bigr )\im \chi_\perp(i\Omega^\varepsilon_{nm}) \notag \\
%% vanish due to the symmetry relation
%%\bigl [w_{r1}(\bm{p})\bigr ]_{n_1m_1}^{\alpha_1\beta_1} \bigl [\bar{w}_{r2}(-\bm{p})\bigr ]_{m_1n_1} ^{\beta_1\alpha_1} =   \bigl [w_{r2}(\bm{p})\bigr ]_{n_1m_1}^{\alpha_1\beta_1} \bigl [\bar{w}_{r1}(-\bm{p})\bigr ]_{m_1n_1} ^{\beta_1\alpha_1} 
%
%
-2\pi T\Gamma^{\rm sf}
\delta^{\alpha_1\alpha_3}
\delta^{\alpha_2\alpha_4}
\bigl(1-\delta_{n_1n_3}\bigr)\delta_{rr^\prime}
\lambda_r
\Bigr [
\delta_{jj^\prime}
\bigl (\delta_{j0}-\delta_{j3}\bigr )
\re {\widehat \chi}(i\Omega_{13}^\varepsilon)
+
\bigl (\delta_{j0}\delta_{j^\prime 3}- \delta_{j3}\delta_{j^\prime 0}\bigr ) 
 i \im{\widehat \chi}(i\Omega_{13}^\varepsilon)
 \Bigr ]
 \notag \\
+2\pi T \delta_{jj^\prime}\delta_{rr^\prime}
\delta^{\alpha_1\alpha_2}
\delta^{\alpha_3\alpha_4}
\Bigr [
\Gamma_j
\bigl ( \delta_{r0}+\delta_{r3}\bigr) 
\delta^{\alpha_2\alpha_3}
+
 \Gamma^{\rm sf} 
\delta_{r3}\bigl (\delta_{j1}+\delta_{j2}\bigr )\re {\widehat \chi}(i\Omega^\varepsilon_{12})
\Bigr ]
\Biggr \} .
\label{eq:GA}
\end{gather}
\end{widetext}
Here we introduced the following notations: $w_{rj} = \spp[w t_{rj}]/4$, where $\spp$ denotes the trace over spin and particle-hole indices, $\lambda_r = \{1,-1,-1,1\}$, $\Omega_{12}^\varepsilon = \varepsilon_{n_1}-\varepsilon_{n_2}$, $\Omega_{13}^\varepsilon = \varepsilon_{n_1}-\varepsilon_{n_3}$,
and $\widehat{\chi}(i\omega)=\chi_+(i\omega)/\chi_+(i0)$.
The diffusion coefficient is given as $D = g/(16 Z_\omega)$. The parameter $\Gamma^{\rm sf} = {n_s(\pi\nu J)^2}S(S+1)\chi(i0)  /{(4\pi )}$ characterizes the strength of interaction due to the inelastic spin-flip scattering. The effective Zeeman splitting for electrons is given as $\tilde{b}_{\rm e} = b_{\rm e} + \pi n_s \nu J \langle S_z\rangle_{\bm S}/(2Z_s)$. The matrices $\mu^{(d)}_{jj^\prime}$ and $\mu^{(c)}_{jj^\prime}$ are defined as follows
\begin{equation}
\mu^{(d)}_{jj^\prime} =
\begin{pmatrix}
0 & 0 &  0 & 0 \\
0 & 0 & 1 & 0 \\
0 & -1 & 0 & 0 \\
0 & 0 & 0 & 0 
\end{pmatrix}_{jj^\prime} ,\quad
\mu^{(c)}_{jj^\prime} =
\begin{pmatrix}
0 & 0 &  0 & 1 \\
0 & 0 & 0 & 0 \\
0 & 0 & 0 & 0 \\
1 & 0 & 0 & 0 
\end{pmatrix}_{jj^\prime}  .
\end{equation}

The second line in Eq. \eqref{eq:GA} involves the elastic spin-flip time $\tau^{\rm sf}_{rj}$. It can be expressed in terms of the static spin susceptibilities as follows
\begin{equation}
\frac{1}{\tau^{\rm sf}_{rj}} =
\frac{1}{\tau^{\rm sf}_\|}
 \zeta_{rj}^\|+ \frac{1}{\tau^{\rm sf}_\perp} \zeta_{rj}^\perp ,
\end{equation}
where  $1/\tau^{\rm sf}_\| = 2 M_2/\tau_{s0}$, 
$1/\tau^{\rm sf}_\perp = T \chi_+(i0)/\tau_{s0}$, and 
\begin{equation}
\zeta_{rj}^\| =\begin{pmatrix}
0 & 1 & 1 & 0 \\
1 & 0 & 0 & 1\\
1 & 0 & 0 & 1\\
0 & 1 & 1 & 0 \\
\end{pmatrix}_{rj}, \quad 
\zeta_{rj}^\perp =\begin{pmatrix}
0 & 1 & 1 & 2 \\
2 & 1 & 1 & 0\\
2 & 1 & 1 & 0\\
0 & 1 & 1 & 2 \\
\end{pmatrix}_{rj} .
\end{equation}
For a sake of convenience,  we note that
\begin{equation}
\chi_+(i0)  = - \frac{2M_1}{b_{\rm i}}
 = \begin{cases}
2/(3T), & \, |b_{\rm i}| \ll T , \\
2/[|b_{\rm i}|(S+1)], & \, |b_{\rm i}| \gg T .
\end{cases}
\label{eq:chi0}
\end{equation}

In the limit of zero Zeeman splitting, $b_i\to 0$, the elastic spin-flip rate becomes equal ${1}/{\tau^{\rm sf,(0)}_{rj}} = {2\zeta_{rj}}/{(3\tau_{s0})}$, where the matrix $\zeta_{rj}$ is defined as follows
\begin{equation}
\zeta_{rj} =\begin{pmatrix}
0 & 2 & 2 & 2 \\
3 & 1 & 1 & 1\\
3 & 1 & 1 & 1\\
0 & 2 & 2 & 2 \\
\end{pmatrix}_{rj} .
\end{equation}
Taking into account that the bare value of the parameter $Z_\omega$ is equal $\pi \nu/4$, we obtain the well-known values for the elastic spin-flip rates in different diffusive modes (see, e.g. Ref. \cite{Efetov1980b}).

The function $h(i\varepsilon_n)$ in the second line of Eq. \eqref{eq:GA} describes the effect of the inelastic scattering off magnetic impurities on the part  of the propagator of the diffusive modes which is diagonal in the Matsubara space. This function is defined as  ($\varepsilon_n>0$):
\begin{gather}
h(i\varepsilon_n) = \sum_{\varepsilon_n>\omega_k>0} \re \widehat{\chi}(i\omega_k) =
\frac{b_{\rm i}}{2\pi T} \im \Bigl [ \psi\Bigl ( 1+
\frac{ib_{\rm i}}{2\pi T}\Bigr )
\notag \\
-\psi\Bigl ( 1+n+
\frac{ib_{\rm i}}{2\pi T}\Bigr )
\Bigr ] . 
\end{gather}
Here $\psi(z)$ denotes the Euler digamma function.
The function $h(i\varepsilon_n)$ appears as the self-energy correction to the diffusive modes in the diagrammatic approach \cite{Ohkawa1983}.
In particular, the function $h$ contains the additional contribution due to inelastic spin-flip on magnetic impurities to decay rate of ``Cooperons'' which has been  studied recently in Ref. \cite{Kashuba2016} in detail. 
In order to discuss this effect, it is convenient to make analytic continuation $i\varepsilon_{n_1} \to \varepsilon_+=\varepsilon+\Omega/2$ and 
$i\varepsilon_{n_2} \to \varepsilon_-=\varepsilon-\Omega/2$. The retarded function $h^R(\varepsilon)$ corresponding to the Matsubara function $h(i\varepsilon_n)$ is given as
\begin{gather}
h^R(\varepsilon) = \frac{b_{\rm i}}{2\pi T} \Biggl [
\frac{1}{2} \sum_{\sigma=\pm} i \sigma \psi\Bigl ( \frac{1}{2} - \frac{i\varepsilon}{2\pi T}+
\frac{ib_{\rm i}\sigma}{2\pi T}\Bigr )
\notag \\
+ \im \psi\Bigl ( 1+
\frac{ib_{\rm i}}{2\pi T}\Bigr )
\Biggr ] .
\end{gather}
The real part of $h^R(\varepsilon)$ determines the additional contribution to the decay rate of the diffusive modes:
\begin{gather}
\frac{1}{\tau_{\rm inel}^{\rm sf}(\varepsilon)} = 
\frac{2}{\tau^{\rm sf}_\perp}  \re h^R(\varepsilon) = 
- \frac{1}{\tau^{\rm sf}_\perp} \Biggl [
1 - \frac{b_{\rm i}}{4T}\Bigl ( 2\coth \frac{b_{\rm i}}{2\pi T}\notag \\
 - \tanh \frac{b_{\rm i}+\varepsilon}{2\pi T}-\tanh \frac{b_{\rm i}-\varepsilon}{2\pi T}\Bigr )
\Biggr ]  .
\end{gather} 
Here we took into account that $\re h^R(\varepsilon)$ is even function of $\varepsilon$. 
%The detailed discussion of the decay rates of ``Cooperons'' in the present case can be found in Ref. \cite{?}. 
Interestingly, the function $h^R(\varepsilon)$ produces also the imaginary correction which is linear in $\Omega$ at $\Omega\to 0$:
\begin{gather}
\frac{1}{\tau^{\rm sf}_\perp} \Bigl [ h^R(\varepsilon_+)+h^R(\varepsilon_-) \Bigr ] 
= \frac{1}{\tau_{\rm inel}^{\rm sf}(\varepsilon)} 
-i (z(\varepsilon) -1) \Omega + \dots .
\label{eq:factor:z}
\end{gather}
Here the frequency renormalization factor is given as follows 
\begin{equation}
z(\varepsilon) = 1 + \gamma^{\rm sf}  \frac{b_{\rm i}}{4\pi T}
\sum_{\sigma=\pm}
 \im \psi^\prime\Bigl ( \frac{1}{2} +
\frac{i(b_{\rm i}+\sigma\varepsilon)}{2\pi T}\Bigr ) ,
\end{equation}
where $\gamma^{\rm sf}  = \Gamma^{\rm sf}/Z_\omega$. 
In the case $|\varepsilon|, T \ll |b_{\rm i}|$ the renormalization factor becomes $z(\varepsilon) = 1 + \gamma^{\rm sf}$ where the parameter  $\gamma^{\rm sf}$ is given as 
$\gamma^{\rm sf} = 1/[\pi (S+1)\tau_{s0} |b_{\rm i}|] \ll 1$ (see Eq. \eqref{eq:chi0}). We note that in the case $|\varepsilon|, T \ll |b_{\rm i}|$ the expansion \eqref{eq:factor:z} holds for $|\Omega| \ll |b_{\rm i}|$.

%%%%%%%%%%%%%%%%%%%%%%%%%%%%%%%%%%%%%%%%%%%%%%%%%%%%%%%%%%%%%%%%%%%%%%%%%%%%
\begin{figure}[t]
\includegraphics[width=0.48\textwidth]{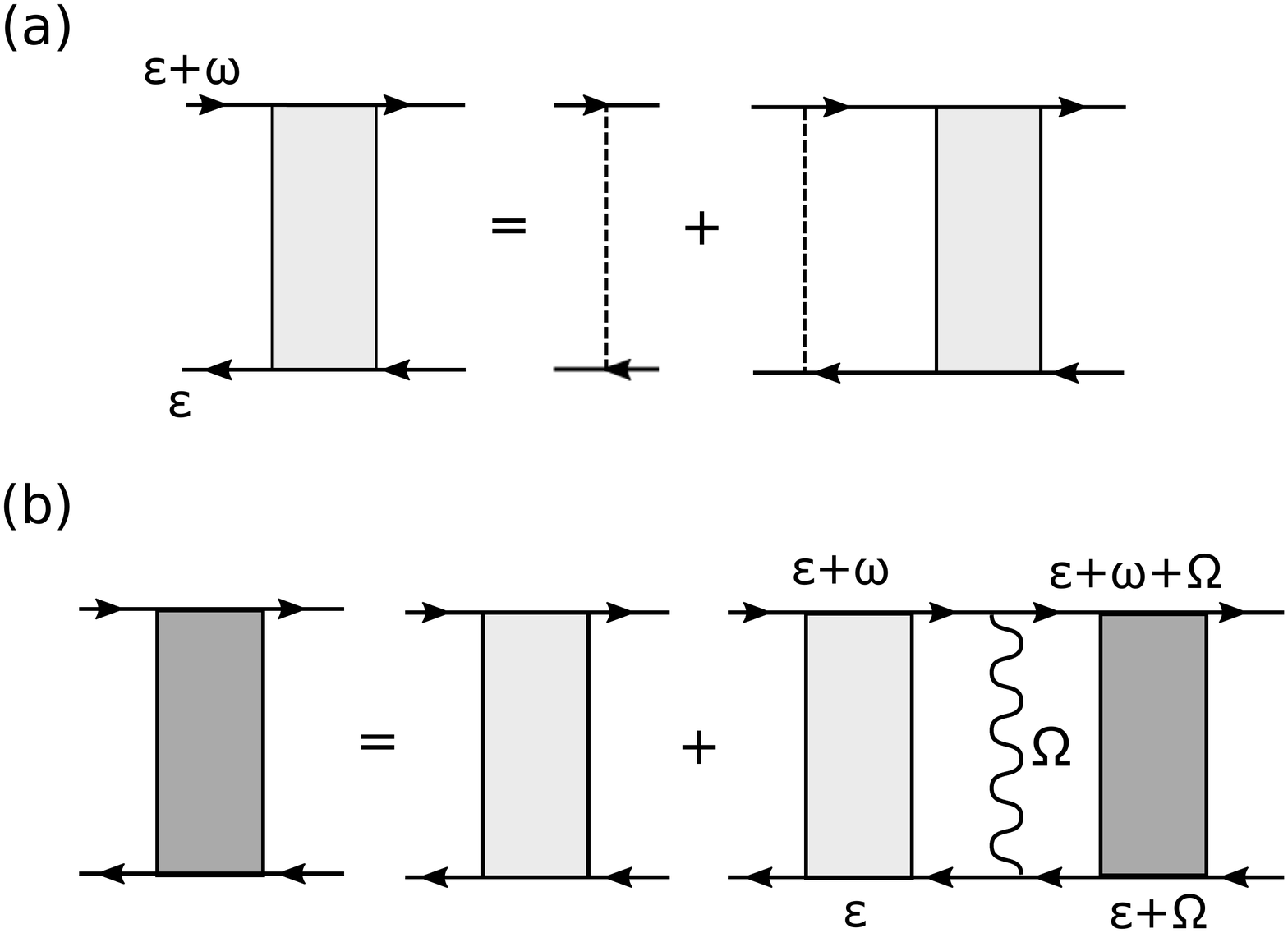}
\caption{(Color online) The diagrammatic representation of equations for the diffusive propagators  $\mathcal{D}_p(i\omega)$ (panel (a)) and 
$\mathcal{D}_p^{(rj)}(i\omega)$ (panel (b)). The solid line stands for the electron Green's function averaged over potential disorder and with self-energy correction due to scattering off magnetic impurities. The dashed line denotes the scattering off the potential disorder. The wavy line stands for the inelastic scattering off magnetic impurities. The Matsubara energies $\varepsilon$, $\omega$ and $\Omega$ are assumed to satisfy the following conditions: $\varepsilon<0$, $\varepsilon+\omega>0$, $\varepsilon+\Omega<0$, and $\varepsilon+\omega+\Omega>0$.
}
\label{Figure1}
%\vspace{-0.5cm}
\end{figure}
%%%%%%%%%%%%%%%%%%%%%%%%%%%%%%%%%%%%%%%%%%%%%%%%%%%%%%%%%%%%%%%%%%%%%%%%%%% 

Since in what follows we are interested in the regime $|b_{\rm e}|\ll T \ll |b_{\rm i}|$ we neglect terms with the spin-flip rates and Zeeman splitting 
in the second line of Eq. \eqref{eq:GA}. Then for frequencies which are much smaller than $|b_{\rm i}|$, we find the following result for the propagators of different diffusive modes:
\begin{widetext}
\begin{gather}
\Bigl \langle [w_{rj}(\bm{p})]^{\alpha_1\alpha_2}_{n_1n_2} [\bar{w}_{r j}(-\bm{p})]^{\alpha_4\alpha_3}_{n_4n_3} \Bigr \rangle =  \frac{2}{g} \delta_{n_{12},n_{34}}  \Biggl \{ \delta^{\alpha_1\alpha_3} \delta^{\alpha_2\alpha_4}  \Biggl [ \delta_{n_1n_3} \mathcal{D}_p(i\Omega_{12}^\varepsilon) - \frac{2 \pi T \gamma_j}{D}\delta^{\alpha_1\alpha_2}  
(\delta_{r0}+\delta_{r3})\mathcal{D}^{(rj)}_p(i\Omega_{12}^\varepsilon)
\widetilde{\mathcal{D}}_p^{(rj)}(i\Omega_{12}^\varepsilon)
\notag \\
+\frac{2 \pi T \gamma^{\rm sf}}{D}
\delta^{\alpha_1\alpha_3} \delta^{\alpha_2\alpha_4}\lambda_r
%\notag\\ \times 
\bigl(\delta_{j0}-\delta_{j3}\bigr )
{\mathcal{D}}_p(i\Omega_{12}^\varepsilon)
{\mathcal{D}}^{(rj)}_p(i\Omega_{12}^\varepsilon)
\Biggr ]
- \frac{2 \pi T \gamma^{\rm sf}}{D}
\delta^{\alpha_1\alpha_2} \delta^{\alpha_3\alpha_4} 
\delta_{r3}\bigl(\delta_{j1}+\delta_{j2}\bigr )
\Bigl [\widetilde{\mathcal{D}}^{(rj)}_p(i\Omega_{12}^\varepsilon)\Bigr ]^2
\Biggr \} 
 .
\label{eq:prop:PH}
\end{gather}
\end{widetext}
Here the following propagator:
\begin{equation}
\bigl [\mathcal{D}_p(i\omega_n)\bigr ]^{-1}
 =  p^2 + (1+\gamma^{\rm sf}) \frac{|\omega_n|}{D} 
\label{eq:freeD}
\end{equation} 
describes ``diffuson'' (for $r=0,3$) and ``cooperon'' (for $r=1,2$) modes in the absence of electron-electron interaction. The factor $1+\gamma^{\rm sf}$ appears as a result of taking into account the self-energy contributions due to scattering off magnetic impurities (see Fig. \ref{Figure1}). The propagator  
\begin{equation}
\bigl [\mathcal{D}^{(rj)}_p(i\omega_n)\bigr ]^{-1}
 =  p^2 +\Bigl [ 1+\gamma^{\rm sf}
 - \gamma^{\rm sf}  \lambda_r (\delta_{j0}-\delta_{j3}) \Bigr ]\frac{|\omega_n|}{D}
\label{eq:freeDD}
\end{equation} 
accounts for the vertex insertions of the scattering off a magnetic impurity into the ``diffuson'' and ``cooperon'' ladder (see Fig. \ref{Figure1}). The electron-electron interaction appears in the propagator of ``diffuson'' modes ($r=0,3$) dressed by electron-electron scattering (see Ref. \cite{Finkelstein1990} for details):
\begin{gather}
\bigl [\widetilde{\mathcal{D}}_p^{(rj)}(i\Omega_{12}^\varepsilon)
 \bigr ]^{-1} =
 \bigl [\mathcal{D}_p^{(rj)}(i\Omega_{12}^\varepsilon)\bigr ]^{-1}
 +\frac{\gamma_j \Omega_{12}^\varepsilon}{D} .
 \label{eq:intD}
 \end{gather}
 For frequencies larger than $|b_{\rm i}|$ the propagators are given by Eq. \eqref{eq:prop:PH} with $\gamma^{\rm sf}$ set to zero.

We note that the form \eqref{eq:prop:PH} of the propagators for the diffusive modes guarantees that the polarization operator is independent of the scattering on magnetic impurities. Indeed, one expects that the self-energy and vertex corrections in polarization bubble due to scattering off a magnetic impurity  cancel each other. 
In order to check it, we write the density-density response (reducible polarization operator with respect to the Coulomb interaction) as follows (see Ref. \cite{Finkelstein1990}):
\begin{align}
\Pi(\bm{q},i\omega_n) = &
- \frac{4}{\pi}(Z_\omega+\Gamma_s) \Bigl [
1 -\pi T (Z_\omega+\Gamma_s) \notag \\
& \times \bigl \langle \Tr I_n^\alpha Q(\bm{q}) \Tr I_{-n}^\alpha Q(-\bm{q}) \bigr \rangle \Bigr ] .
\end{align}
Evaluation of the average with the help of Eq. \eqref{eq:prop:PH} in the lowest order with respect to $1/g$ (this approximation corresponds to the random phase approximation) results in the following form of the polarization operator:
\begin{equation}
 \Pi^{\rm RPA} (\bm{q},i\omega_n) =
 - \frac{4}{\pi}  \frac{Z_\omega (1+\gamma_s) D q^2}{Dq^2+(1+\gamma_s) |\omega_n|} .
\end{equation} 
As expected, the parameter $\gamma^{\rm sf}$  drops from the 
expression for $\Pi^{\rm RPA}$ due to cancelation of self-energy and vertex contributions.

\section{Evaluation of the conductivity\label{Sec:Cond}}

\subsection{Kubo formula}

Within the formalism of the nonlinear sigma model the static conductivity can be computed by means of the following Kubo formula:
\begin{align}
g^\prime = & -\frac{g}{16 n} \Bigl \langle \Tr [J_n^\alpha,Q(\bm{r})] [J_{-n}^\alpha,Q(\bm{r})] \Bigr \rangle +\frac{g^2}{64 d n} \int d\bm{r}^\prime  \notag \\
& \times 
\Bigl \langle \Tr J_n^\alpha Q(\bm{r}) \nabla Q(\bm{r})
\Tr J_{-n}^\alpha Q(\bm{r}^\prime) \nabla Q(\bm{r}^\prime) \Bigr \rangle ,
\label{eq:PO:g}
\end{align}
where $d$ stands for dimensionality, the limit $n \to 0$ is assumed, and 
\begin{equation}
J_n^\alpha = \frac{t_{30}-t_{00}}{2} I_n^\alpha + \frac{t_{30}+t_{00}}{2} I_{-n}^\alpha .
\end{equation}
The average $\langle\dots\rangle$ in Eq. \eqref{eq:PO:g} is defined with respect to the total action $\mathcal{S}_\sigma+\mathcal{S}_\textrm{mag}$. Evaluating the averages in Eq. \eqref{eq:PO:g} with the help of  Eq. \eqref{eq:prop:PH}, we find that the conductance in the one-loop approximation  can be written as follows
\begin{equation}
g^\prime = g +\delta g^{\rm wl} + \delta g^{\rm AA} +\delta g^{\rm sf}_1+ \delta g^{\rm sf}_2  .
\label{eq:dg:1loop}
\end{equation}
Here $\delta g^{\rm wl}$ represents the interference correction. It has the standard form \cite{Wegner1979,Gorkov1979,Efetov1980b}:
\begin{equation}
\delta g^{\rm wl} = 
\sum_{r=1,2} \sum_j \bigl (2\delta_{j0}-1\bigr )
\int \frac{d\bm{p}}{(2\pi)^d}\mathcal{D}^{(rj)}_p(0) .
\label{eq:dg:wl}
\end{equation}
Since the weak-localization correction involves ``cooperon'' 
modes at zero frequency, the spin-flip scattering affects $\delta g^{\rm wl}$ only via decay rate of ``cooperon'' modes (see Ref. \cite{Kashuba2016} for detailed discussion). 

\begin{figure}[t]
\includegraphics[width=0.48\textwidth]{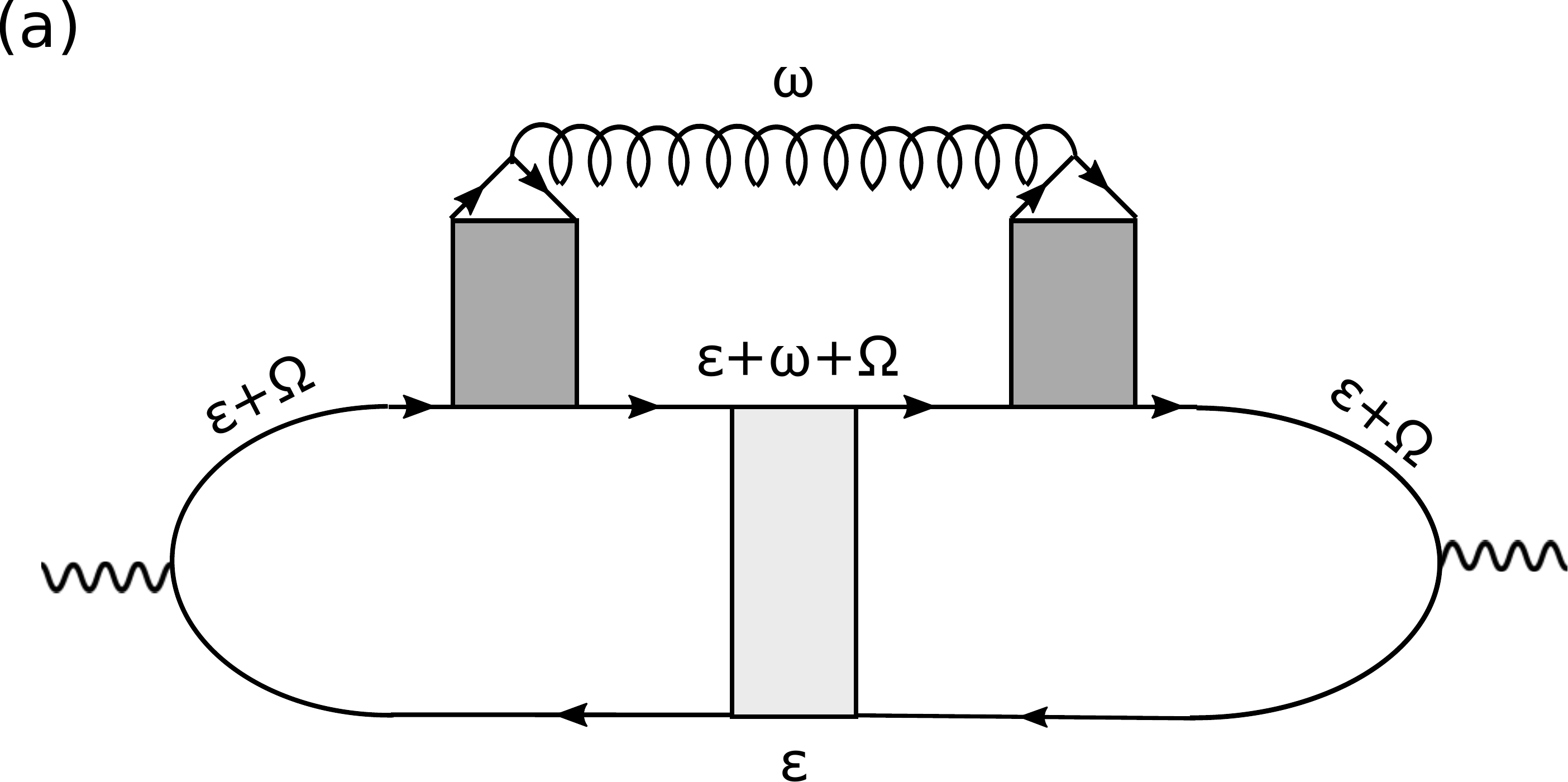}
\includegraphics[width=0.48\textwidth]{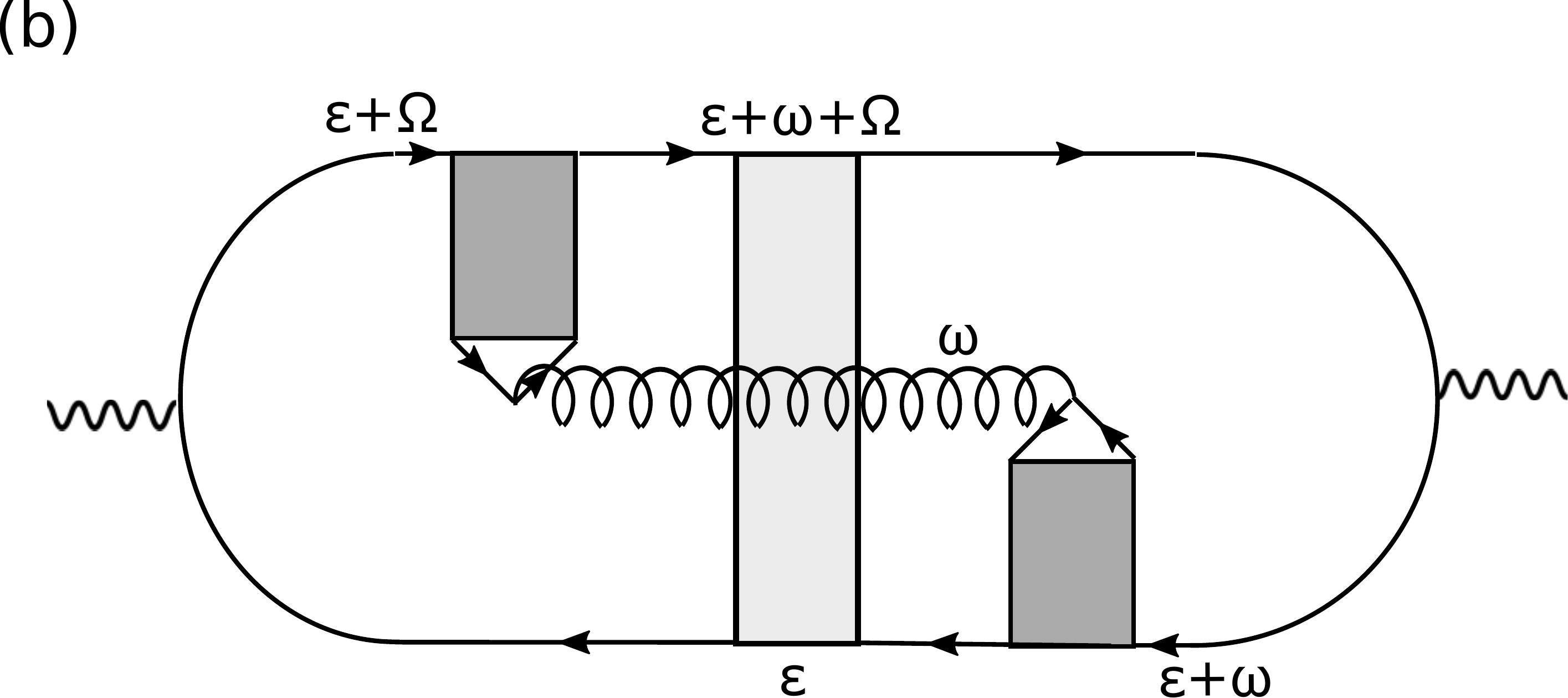}
\caption{(Color online) The sketch of diagrams contributing to the 
Altshuler-Aronov corrections. The spring-like line stands for the electron-electron interaction. The other elements have the same meaning as in the previous figure.
}
\label{Figure2}
%\vspace{-0.5cm}
\end{figure}

Next term, $\delta g^{\rm AA}$, in the right hand side of Eq. \eqref{eq:dg:1loop} is the Altshuler-Aronov correction due to electron-electron interaction \cite{Altshuler1979b,Finkelstein1983a,Finkelstein1983b}:
\begin{align}
\delta g^{\rm AA}= & \frac{128 \pi T}{n g d} \sum_{r=0,3}\sum_{j} \Gamma_j
\int \frac{d\bm{p}}{(2\pi)^d} \, p^2
\sum_{m>0}  
 \min\{m,n\} \notag \\
& \times \mathcal{D}^{(rj)}_p(i\omega_m)\widetilde{\mathcal{D}}^{(rj)}_p(i\omega_m)
\mathcal{D}_p(i\omega_{m+n})\ .
\end{align}
Here the limit $n\to 0$ is assumed. We emphasize that for $|\omega_m|\ll |b_{\rm i}|$ the spin-flip scattering \emph{does} enter the expression for $\delta g^{\rm AA}$ via the frequency renormalization factors in the diffusion propagator. We mention that the Altshuler-Aronov correction involves three types of propagators of diffusive modes (see Fig. \ref{Figure2}).

Performing analytic continuation to the real frequencies, $i\omega_n \to \omega +i 0$, and taking the limit $\omega\to 0$ we obtain the following result
\begin{align}
\delta g^{\rm AA}= &  \frac{64}{g d} \im \sum_{j} \Gamma_j
\int \frac{d\bm{p}}{(2\pi)^d} \, p^2
\int d\Omega \, \partial_\Omega \Bigl (\Omega \coth \frac{\Omega}{2T}\Bigr ) \notag \\
& \times \mathcal{D}^{R}_p(\Omega)  \mathcal{D}^{(0j),R}_p(\Omega) \widetilde{\mathcal{D}}^{(0j),R}_p(\Omega) .
\label{eq:AA:R}
\end{align}
Here we took into account that diffusion propagators with $r=0$ and $r=3$ coincide. The propagators $\mathcal{D}^R_p(\Omega)$, $\mathcal{D}^{(rj),R}_p(\Omega)$, and $\widetilde{\mathcal{D}}^{(rj),R}_p(\Omega)$ denote for the retarded propagators corresponding to $\mathcal{D}_p(i\Omega)$, $\mathcal{D}^{(rj)}_p(i\Omega)$, and $\widetilde{\mathcal{D}}^{(rj)}_p(i\Omega)$, respectively.

%%%%%%%%%%%%%%%%%%%%%%%%%%%%%%%%%%%%%%%%%%%%%%%%%%%%%%%%%%%%%%%%%%%%%%%%%%%%
%%%%%%%%%%%%%%%%%%%%%%%%%%%%%%%%%%%%%%%%%%%%%%%%%%%%%%%%%%%%%%%%%%%%%%%%%%% 

Next there is the following correction due to inelastic spin-flip scattering: 
\begin{gather}
 \delta g^{\rm sf}_1 =  - \frac{64 \pi T \Gamma^{\rm sf}}{n g}
 \sum_{j=0,3}  (-1)^j
 \int \frac{d\bm{p}}{(2\pi)^d} \sum_{m>0} m\,
 {\mathcal{D}}_p(i\omega_{m+n}) \notag \\
 \times 
  \sum_{r=0,3} {\mathcal{D}}^{(rj)}_p(i\omega_{m+n})
\Biggl [
1 - p^2 \Bigl [{\mathcal{D}}_p(i\omega_{m})\notag \\
+{\mathcal{D}}_p(i\omega_{m+2n})\Bigr ] \Biggr ]  .
 \end{gather} 
We remind a reader that the limit $n\to 0$ is assumed. 
On the first glance, it seems that this limit is not finite such that correction $\delta g^{\rm sf}_1$ violates the gauge invariance. However, taking into account that the diffusion propagators 
 $\mathcal{D}^{(r j)}$ are the same for $r=0$ and $r=3$, 
 we can rewrite this correction $\delta g^{\rm sf}_1$ 
as the sum of two corrections, $\delta g^{\rm sf}_1=
\delta g^{\rm sf}_{1,\omega}+\delta g^{\rm sf}_{1,f}$, where $\delta g^{\rm sf}_{1,\omega}$ seems to have no finite limit at $n\to 0$ and $\delta g^{\rm sf}_{1,f}$ has a smooth $n\to 0$ limit. In particular, we find 
\begin{equation}
\delta g^{\rm sf}_{1,\omega} =   \frac{2}{d n}\sum_{j=0,3}  \sum_{m>0} \int \frac{d\bm{p}}{(2\pi)^d} 
\frac{\partial}{\partial \bm{p}} \frac{\partial}{\partial \bm{p}}
  \ln \frac{\mathcal{D}_p(i\omega_{m})}{\mathcal{D}^{(0j)}_p(i\omega_{m})} .
\end{equation} 
Since  $\delta g^{\rm sf}_{1,\omega}$ 
has the form of the second derivative
with respect to the momentum  this correction is determined by the ultra-violet of the low-energy effective theory. Therefore, we cannot accurately compute it within the nonlinear sigma model approach. However this correction is second order in $\gamma^{\rm sf}$ so taking it into account is accuracy excess.

After analytic continuation to the real frequencies, $i\omega_n \to \omega +i 0$, and taking the limit $\omega\to 0$ the finite correction $\delta g^{\rm sf}_{1,f}$ can be written as
\begin{align}
\delta g^{\rm sf}_{1,f} = &
\frac{1}{d} \re \sum_{j=0,3} 
\int \frac{d\bm{p}}{(2\pi)^d} \, p^2
\int d\Omega \, \partial_\Omega \Bigl (\coth \frac{\Omega}{2T}\Bigr ) \notag \\
& \times \Bigl [\mathcal{D}^{R}_p(\Omega)-  \mathcal{D}^{(0j),R}_p(\Omega) \Bigr ]^2 .
\label{nonlog}
\end{align}

The last correction in Eq. \eqref{eq:dg:1loop} is also due to the inelastic spin-flip scattering represented by the last term in Eq. \eqref{eq:prop:PH}. It has the following form:
\begin{align}
\delta g^{\rm sf}_2=  & \frac{128 \pi T \Gamma^{\rm sf}}{n g d} \sum_{j=1,2} 
\int \frac{d\bm{p}}{(2\pi)^d} \, p^2
\sum_{m>0}  
 \min\{m,n\} \notag \\
& \times
\Bigl [\widetilde{\mathcal{D}}^{(3j)}_p(i\omega_m)\Bigr ]^2
\mathcal{D}_p(i\omega_{m+n}) .
\label{eq:dg:sf:fin0}
\end{align} 
 Here, again, the limit $n\to 0$ is assumed. Diagrammatically, this correction has the structure similar to diagrams shown in Fig. \ref{Figure2} in which the electron-electron interaction line should be substituted by the dynamical spin susceptibility. Performing analytic continuation to the real frequencies, $i\omega_n \to \omega +i 0$, and taking the limit $\omega\to 0$, we obtain the following result
\begin{align}
\delta g^{\rm sf}_2=  & \frac{64 \Gamma^{\rm sf}}{g d} \im
\int \frac{d\bm{p}}{(2\pi)^d} \, p^2
\int d\Omega\,\, \partial_\Omega \Bigl ( \Omega \coth \frac{\Omega}{2T}\Bigr )
\notag \\
& \times \Bigl [\widetilde{\mathcal{D}}^{(31),R}_p(\Omega)\Bigr ]^2
\mathcal{D}^{R}_p(\Omega) .
\label{eq:dg:sf:fin}
\end{align} 
Here we took into account the equivalence of diffusion propagators with $j=1$ and $j=2$.  
It is worthwhile to mention that the correction  $\delta g^{\rm sf}$ involves triplet diffusive modes with the total spin projection equal $\pm 1$. We note that the correction \eqref{eq:dg:sf:fin} is similar to the quantum correction due to electron-paramagnon scattering \cite{Paul}.

\subsection{Logarithmic corrections to conductance due to inelastic spin-flip scattering} 
 
As we mentioned above, in this paper we focus on the case $T \ll |b_i|$. Also we are interested in corrections of the second order in $J$ and in two-dimensional case. Then, expanding the correction \eqref{eq:AA:R} to the first order in $\gamma^{\rm sf}$, we find 
\begin{align}
\delta g^{\rm AA} = & -\frac{1}{\pi} \sum_{j=0}^3 \Bigl [1 - \frac{1+\gamma_j}{\gamma_j} \ln(1+\gamma_j) \Bigr ] \ln \frac{1}{2\pi T\tau}
\notag \\
& - \frac{\gamma^{\rm sf}}{\pi} 
\Bigl [ \frac{1}{2} + \frac{1}{\gamma_s}- \frac{1+\gamma_s}{\gamma_s^2}\ln(1+\gamma_s)\Bigr ] \ln  \frac{|b_{\rm i}|}{2\pi T}
\notag \\
&
- \frac{2\gamma^{\rm sf}}{\pi} 
\Bigl [ 1- \frac{1}{\gamma_t}  \ln(1+\gamma_t) \Bigr ] \ln  \frac{|b_{\rm i}|}{2\pi T} \notag \\
& 
- \frac{\gamma^{\rm sf}}{\pi} 
\Bigl [ \frac{3}{2} - \frac{1}{\gamma_t}+ \frac{1-\gamma_t}{\gamma_t^2}\ln(1+\gamma_t)\Bigr ] \ln  \frac{|b_{\rm i}|}{2\pi T}
 .
 \label{eq:AA:R:res} 
\end{align}
Here the first line represents the standard Altshuler-Aronov correction to the conductivity. Since the corresponding contribution exists for frequencies larger then $|b_{\rm i}|$ the ultra-violet cutt-off for this correction is inverse transport mean free time $1/\tau$. The second line describes the correction due to the effect of the inelastic scattering off magnetic impurities on the singlet particle-hole channel. The third line corresponds to the correction from triplet particle-hole channel with the total spin projection equal $\pm 1$. The forth line describes the correction from the triplet particle-hole channel with the zero total spin projection.
We note that the corrections proportional to $\gamma^{\rm sf}$  involve $\ln ({|b_{\rm i}|}/{2\pi T})$ and vanish in the absence of electron-electron interaction. We mention that 
in the standard Altshuler-Aronov correction  (the first line of 
Eq. \eqref{eq:AA:R:res}) the singlet channel favours localization (since $\gamma_s \leqslant 0$) whereas the triplet channel favours antilocalization (since $\gamma_t \geqslant 0$) at low temperature. The corrections proportional to $\gamma^{\rm sf}$ work in the opposite direction, i.e. the presence of inelastic scattering off magnetic impurities decreases the effect of localization (antilocalization) in the singlet (triplet) channels, respectively.

The correction \eqref{nonlog} does not produce logarithmic terms since the integral over frequencies is restricted by $|\Omega| \lesssim T$. The other correction due to inelastic scattering, Eq. \eqref{eq:dg:sf:fin}, reads   
 \begin{align}
\delta g^{\rm sf}_2 = 
\frac{\gamma^{\rm sf}}{\pi \gamma_t}
\Bigl [ 1 -\frac{1}{\gamma_t} \ln (1+\gamma_t)\Bigr ] \ln \frac{|b_{\rm i}|}{2\pi T} .
\label{eq:dg:sf:fin:fin}
\end{align} 
We note that this correction is positive, i.e. works in favour of antilocalization at low temperatures.  
In the absence of electron-electron interaction $\delta g^{\rm sf}_2$ is the only correction to the conductivity due to the inelastic scattering off magnetic impurities. It acquires the following form:
\begin{equation}
\delta g^{\rm sf}_2  \to  \frac{\gamma^{\rm sf}}{2} \ln \frac{|b_{\rm i}|}{2\pi T} =\frac{1}{2\pi (S+1)\tau_{s0} |b_{\rm i}|} \ln \frac{|b_{\rm i}|}{2\pi T}\ .
\end{equation}
This quantum correction works in opposite direction with respect to the weak-localization correction.

\section{Discussions and conclusions \label{Sec:Sum}}

The temperature dependent corrections to the conductivity discussed above were derived within the Born approximation for scattering off magnetic impurities. We remind that standard Kondo correction to the conductance in the clean system appears beyond Born approximation: in the third order in the exchange interaction. In the case $T \ll |b_{\rm i}|$ this correction is temperature independent since the infrared cut-off for the Kondo logarithm is given by $|b_{\rm i}|$ rather than $T$. In disordered case, the inelastic corrections to the conductance studied previously \cite{Ohkawa1983,Ohkawa1984,Suga1986} has been also of the third order in the exchange interaction. Since the corrections \eqref{eq:AA:R:res} and \eqref{eq:dg:sf:fin:fin} are of the second order in the exchange interaction they are more important for small enough $\nu J$. We note that the corrections of the third order in $J$ for the case $|b_{\rm e}| \ll T \ll |b_{\rm i}|$ have not been computed yet. Therefore, we cannot compare the second and third order corrections quantitatively.

In the absence of electron-electron interaction the structure of the correction \eqref{eq:dg:sf:fin} is similar to correction to the conductivity of disordered electron system in diffusive regime due to electron-electron interaction mediated by inelastic scattering off paramagnons  \cite{Paul}. Difference between paramagnons and magnetic impurities is in the form of the induced electron-electron interaction. In the latter case, it  is short-ranged and is independent of the transferred frequency for small frequencies. Away from the ferromagnetic quantum phase transition point the induced electron-electron interaction due to paramagnons becomes also momentum and frequency independent, and, consequently, results in the logarithmic-in-$T$ correction to the conductivity in two dimensions \cite{Paul}.

In this paper we consider the case of the electron system in the absence of  spin-orbit splitting. If the spin-orbit splitting is present then it will cut off the diffusion poles of triplet diffusons. Therefore, this results in suppression of the temperature dependence of the correction \eqref{eq:dg:sf:fin:fin} and the contributions which involve $\gamma_t$ in Eq. \eqref{eq:AA:R:res}. The only temperature dependent contribution due to inelastic scattering off magnetic impurities which remains in the case of spin-orbit coupling is the term in the second line of Eq. \eqref{eq:AA:R:res}
which describes modification of the Altshuler-Aronov correction in the singlet channel.

Experimentally, the influence of magnetic impurities on the weak localization correction via the dephasing time induced by the spin-flip scattering has been intensively studied in two-dimensional electron systems for many decades starting from seminal papers \cite{Bergmann1987,Van1987}.  
We are not aware of any systematic experimental studies of the effect of magnetic impurities on the Altshuler-Aronov correction to the conductivity in two-dimensional electron systems. In general, 
clear  separation of the interference and interaction corrections is a difficult experimental problem (see for example, recent papers \cite{Minkov2009,Minkov2011,Minkov2012}). The effects described in the present paper obviously complicate this formidable task.

We note that potential scattering affects also the spin susceptibility of a magnetic impurity resulting in additional (with respect to usual Kondo renormalization) temperature dependent corrections \cite{Ohkawa1983,Zyuzin1984,Suga1988}. Therefore, it would be interesting to consider the corrections to the spin susceptibility of a magnetic impurity in the case of different g-factors and in the presence of electron-electron interaction.

To summarize, we studied the quantum corrections to the conductivity of the two-dimensional disordered interacting electron system in the diffusive regime 
due to inelastic scattering off magnetic impurities. Contrary to previous works, (i) we considered the case of different g-factors for electrons and magnetic impurities, $|g_{\rm e}|  \ll |g_{\rm i}|$; (ii) we focused on the intermediate temperature range $|g_{\rm e}| \mu_B H \ll T \ll |g_{\rm i}| \mu_B H$; (iii) we took into account electron-electron interaction in the particle-hole channel. We found that within the Born approximation the inelastic scattering off magnetic impurities results in additional temperature-dependent correction to the conductivity (cf. Eq. \eqref{eq:dg:sf:fin:fin}). Also the inelastic scattering modifies the Altshuler-Aronov corrections to the conductivity (cf. Eq.\eqref{eq:AA:R:res}). Our predictions present a challenge for experimental studies of low temperature transport in electron disordered systems with rare magnetic impurities.

\begin{acknowledgements}
We thank I. Gornyi for very useful discussions. The work was partially supported by the Alexander von Humboldt Foundation, the Russian Foundation for Basic Research under the Grant No. 17-02-00541, the Ministry of Education and Science of the Russian Federation under the Grant No. 14.Y26.31.0007, and the Basic research program of HSE.\color{black}
\end{acknowledgements}

%%%%%%%%%%%%%%%%%%
\bibliography{biblio}

\end{document}